\documentstyle[preprint,eqsecnum,aps]{revtex}

\begin{document}
\draft
\title{Lifetime of Two-Dimensional Electrons Measured by
Tunneling Spectroscopy}
\author{S.Q. Murphy$^*$, J.P. Eisenstein, L.N. Pfeiffer and K.W. West}
\address{AT\&T Bell Laboratories, Murray Hill, NJ 07974}
\date{\today}
\maketitle
\begin{abstract}
For electrons tunneling between parallel two-dimensional electron
systems, conservation of in-plane momentum produces sharply resonant
current-voltage characteristics and provides a uniquely sensitive
probe of the underlying electronic spectral functions.  We report
here the application of this technique to accurate measurements of
the temperature dependence of the electron-electron scattering rate
in clean two-dimensional systems.  Our results are in qualitative
agreement with existing calculations.
\end{abstract}

%\narrowtext
Several scattering mechanisms contribute to limiting the
quantum lifetime of electrons in semiconductors.  At low
temperature, the lifetime of electrons close to the Fermi
surface is dominated by elastic scattering off of the static
disorder potential in the material.  At higher temperatures,
inelastic processes like electron-phonon and electron-electron
scattering take over.  Electrical transport measurements are often
adequate for the study of the elastic scattering and the
electron-phonon processes. The electron-electron scattering time
$\tau_{ee}$, however, is much harder to extract from transport
experiments since such processes conserve the total momentum of the
electron system.  More sophisticated techniques involving
quantum interference have been used[1] to extract a ``dephasing''
time $\tau_{\phi}$ which is closely related to $\tau_{ee}$.
Though roughly comparable in magnitude, these two times are
not precisely equivalent[2]. In this paper we report a {\it direct}
determination of the lifetime of two-dimensional (2D) electrons
based upon the method of tunneling spectroscopy[3]. While this[4]
and other[5] methods have been applied to determine {\it hot}
electron-LO phonon scattering times, here we are concerned with the
much weaker scattering processes which limit the lifetime of
thermal electrons near the Fermi level.  Crucial to the success of
our method is the conservation of in-plane momentum. This
constraint greatly restricts the phase space available for
tunneling between parallel 2D systems and allows unique access to
the underlying electronic spectral function $A(E,k)$.  This function,
which gives the probability that an electron with wavevector $k$ has
energy $E$, possesses a strong peak near the single-particle energy
$\hbar^2 k^2 /2m$.  The width of this peak, the quantity
measured in these experiments, reflects the finite lifetime of the
momentum eigenstates.

In these experiments we measure the tunnel current flowing
perpendicularly between two parallel 2D electron systems
(2DES) separated by a barrier.  In the ideal case (i.e. no
disorder, electron-electron interactions, etc.) the conservation of
in-plane momentum implies that an electron can
tunnel only if the quantized energy levels in the two wells line
up precisely[6].  This implies that the current-voltage
characteristics of an ideal 2D-2D tunnel junction is singular; the tunneling
conductance is zero everywhere except at
those discrete voltages which produce alignment of the
proper energy levels.  This unusual situation contrasts sharply
with 2D-3D and 3D-3D junctions where tunneling proceeds over an
energy range comparable to the Fermi energy $E_F$.  In real
2D-2D junctions the tunnel resonances will have a finite width, set
not by $E_F$, but only by the degree to which the conditions of
the ideal model break down.  Loss of momentum conservation due to
imperfections in the tunnel barrier is one source of broadening.
But even with a perfect barrier, the finite lifetime of individual
electronic (momentum) states in either 2DES will broaden the tunnel
resonance[7].  As we shall see, the contribution of
electron-electron scattering to this latter effect dominates the
temperature dependence of the observed tunneling linewidth.

The double quantum well (DQW) heterostructures used in this
experiment consist of two $200 \AA$-wide GaAs quantum wells
separated by an undoped ${\rm Al}_x {\rm Ga}_{1-x} {\rm As}$ barrier.
Samples with barrier widths ranging from 175 to $340 \AA$ and
Al mole fractions $0.1<x<0.33$ have been studied.
Total densities ranged from 1 to $1.6 \times 10^{11} cm^{-2}$ with
low temperature mobilities in excess of $10^6 cm^2
/Vs$. The densities in each quantum well could be further adjusted
using gate electrodes deposited on the top and bottom of the
sample.
Ohmic contacts to the individual 2DES layers[8] were
placed at the ends of narrow arms protruding from a $250 \mu m$
square central mesa.
With these contacts we could directly measure
the tunneling conductance $dI/dV$ (using 17Hz, 0.1mV excitation) as
a function of the dc interlayer voltage $V$.

Figure 1 shows typical $dI/dV$ {\it vs.}  $V$ tunnel resonances at four
temperatures between T=0.7 and 10K from a sample with equal 2DES
densities ($N_s =1.6 \times 10^{11} cm^{-2}$) in each
quantum well.  These resonances are centered at zero voltage since,
with equal densities in the two wells, the alignment of the lowest
subband energy levels required for tunneling occurs simultaneously
with the alignment of the Fermi levels.
Note that
the observed widths of the tunnel resonances ($\sim 0.5meV$) are
much less than the Fermi energies ($E_F \approx 5.7meV$) of the
2DES's.  This implies a high degree of momentum conservation on
tunneling[9].
Figure 1 also reveals that as the temperature
rises, the peak tunneling conductance falls while the width of the
resonance increases.  We shall argue that this behavior reflects
the decreasing electronic lifetime, due to electron-electron
scattering, in each quantum well.

Instead of analyzing the measured $dI/dV$ data directly, it is more
convenient to study the ratio $I/V \equiv F(V)$. We construct this
ratio after numerically integrating the measured $dI/dV$ to obtain
the tunnel current[10].  This ratio has two important attributes.
First, as shown below, for momentum conserving tunneling $F(V)$ is
just the convolution of the fundamental spectral functions $A(E,k)$
of the two 2D electron systems.  Second, since $F(V)$ is the
{\it ratio}  of current to voltage, it gives a fair assessment of
the tunneling current which is observed well away from the main
resonance itself. (This non-resonant current is only weakly voltage
dependent and thus hardly appears in $dI/dV$.)  The dotted curves in
the inset to Fig. 2 are the $F(V)$ functions derived from the raw
tunneling data shown in Fig. 1.  As expected, $F(V)$ is strongly
peaked around $V=0$ and falls steadily toward zero off resonance.
The solid curves in the inset are simple Lorentzian fits to the
$F(V)$ data. Only the amplitude and width of the Lorentzian is
adjusted; the fits contain {\it no}  vertical offsets.  It is thus
clear that the tunneling in these samples is dominated by a single,
highly momentum conserving, resonant process and that the
off-resonance tunnel currents are of secondary importance.

Figure 2 also shows the temperature development, for three samples,
of the resonance width $\Gamma$.  The numerical value of $\Gamma$ is
taken to be the half width at half maximum (HWHM) of the
experimental $F(V)$ function. (In spite of the near-Lorentzian shape
of the resonances, lineshape fitting is not employed.) Below about
T=2K $\Gamma$ is temperature independent indicating that inelastic
processes have become negligible.  In this regime the resonance
width is sensitive to density inhomogeneities, momentum
non-conservation due to barrier imperfections, and the finite
lifetime of the electrons in each 2DES produced by scattering off
of the static disorder potential (e.g the Si donors).  For sample A
the low temperature width $\Gamma_0 =0.22meV$ compares favorably
with the quantum lifetime extracted from analysis of the measured
Shubnikov-deHaas resistivity oscillations ($\hbar / \tau
=0.17meV$). This suggests that even at low temperatures the
tunneling resonance width is dominated by the quantum lifetime of
the 2D electrons and not by breakdown of momentum conservation.  We
do not emphasize this point however since it is the finite
temperature linewidth which we are most interested in here.

Figure 2 shows that as the temperature is increased $\Gamma$ grows,
essentially as $T^2$.  This broadening is {\it not}  simply due
to thermal smearing of the Fermi distribution of the 2DES since the
constraint of momentum conservation is indifferent to the thermal
population of the various momentum states. Rather, the increasing
resonance width signals the onset of inelastic processes which
shorten the electronic lifetime. Of the three samples displayed in
Fig. 2 two (A and B) have different levels of disorder, as
evidenced by their different low temperature linewidths $\Gamma_0$, but equal
2DES densities.  Sample C, however, has the same
$\Gamma_ 0$ as sample B but lower density. From this we conclude
that the temperature dependent part of the linewidth depends upon
density but not disorder.  This suggests an inelastic process, like
electron-electron (e-e) or electron-acoustic phonon (e-ph)
scattering.  The e-ph scattering rate can be independently
determined from the temperature dependence of the 2DES mobility
$\mu$. Above a few Kelvin e-ph scattering is not limited to small
angles, and we can directly compare the mobility lifetime
broadening $\hbar / \tau_{\mu}$ to the measured tunnel resonance
linewidth[11].  This comparison shows that $\hbar / \tau_{eph}$ is
roughly 50X smaller than $\Gamma$ at $T=10K$.  Thus e-ph scattering
within each 2DES is a minor contributor to the net electronic
lifetime in this experiment.  Phonon-assisted tunneling processes
can also be ruled out since the opening up of such a new tunneling
channel would presumably increase the peak tunnel conductance
whereas it is observed in fact to {\it decrease} .  We believe
instead that e-e scattering is responsible for the observed
temperature dependence of $\Gamma$.

To quantitatively analyze our data we begin with the generalized
Golden Rule expression[12] for the tunnel current flowing between
the left (L) and right (R) quantum wells:
\begin{eqnarray}
I  = \alpha \sum_{k,k'} |T_{k,k'}|^2 \!\!\!\!&\!\!\!\!\! &\!\!\!\! \int_{-
\infty}^{\infty}\!\!\! dE
\int_{- \infty}^{\infty} \! \! \! dE'
 A_L (E,k) A_R (E',k' ) \nonumber \\
& \times &  [f(E)-f(E')]\: \delta (E-E'-eV)
\end{eqnarray}
where $\alpha$ is a constant, $T_{k,k'}$ the tunneling
matrix element, and {\it f}(E) is the Fermi function.   With $E$ measured
from the Fermi level $E_F$ the spectral function $A(E,k)$ is a
function of $x=E+E_F - \hbar^2 k^2 /2m$ possessing a
sharp peak near[13] x=0.  Near this ``quasiparticle'' peak $A(E,k)$ is
usually taken to be a Lorentzian: $A(x)=( \gamma /2 \pi )/(x^2
+ \gamma^2 /4)$, with $\gamma$ representing the lifetime
broadening $\gamma = \hbar / \tau$ of the quasiparticles. (Note that
the HWHM of $A(E,k)$ is $\gamma /2$.)  Imposing momentum conservation
(i.e. $|T_{k,k'} |^2 = |t|^2 \delta_{k,k'}$) and assuming, for simplicity, that
$k_B T$ and
$\gamma$ are much less than $E_F$ (as in our experiment), reduces
Eq. 1 to
\begin{equation}
{I \over {V}} = \beta |t|^2
\!\! \int_{- \infty}^{\infty}\!\!\! dx A_L (x)
A_R (x+E_{F,R} -E_{F,L} -eV)
\end{equation}
with $\beta$ a constant. Thus, the ratio $I/V \equiv F(V)$ is just the
convolution
of the spectral functions[14] and exhibits a maximum when $eV=E_{F,R} -
 E_{F,L}$, i.e. when the energy levels in the quantum wells
are aligned. If the two layers have the same density this occurs at
$V=0$.  Note that the Fermi functions do not enter in $F(V)$.  As
long as $k_B T$ and $\gamma_{L,R}$ remain much less $E_F$,
this consequence of momentum conservation implies that the observed
temperature dependence of the tunnel resonance comes only from the
spectral widths $\gamma_{L,R}$.  Note also that for Lorentzian
spectral functions, $F(V)$ is also Lorentzian, exhibiting a HWHM of
$\Gamma = ( \gamma_L + \gamma_R )/2$.  Thus, $\Gamma$, our
experimentally measured quantity, is just the average lifetime
broadening $\hbar / \tau$ of an electron in the double layer system.

Figure 3 summarizes our results for the temperature dependence of
$\Gamma$ extracted from tunnel resonances observed with equal 2D
densities in the quantum wells.  In all cases we have found $\Gamma$
to be well approximated by $\Gamma (T)= \Gamma_0 + \alpha T^2$.  The inset to
Fig. 3 shows that the coefficient $\alpha$ is
inversely proportional to the 2D density $N_s$.
Data from
all samples and densities[15] are displayed together in the main panel
of Fig. 3.  The data collapse reasonably well onto a single curve
if, for each density, we plot $( \Gamma - \Gamma_0 )/E_F$
{\it vs.}  $T/T_F$.  This behavior is consistent with
electron-electron scattering in a clean 2DES.

There is disagreement among the various theoretical
calculations of the thermal e-e scattering rate.  Hodges,
{\it et al.} [16] first showed that for a 2D electron at
the Fermi level $\hbar / \tau_{ee} \propto T^2 \log(T_F /T)$ at
low temperature. Subsequently, Giuliani and Quinn (GQ)[17] found
\begin{equation}
{{\hbar} \over {\tau_{ee}}}= {{E_F} \over {2 \pi}}
{\left( {T} \over {T_F} \right)^2}
\left[ \log {\left( {T_F} \over{T} \right) } +1+\log{ \left( {2q_{TF}} \over
{k_F}
\right) } \right]
\end{equation}
where $q_{TF}$ is the 2D Thomas-Fermi screening wavevector ($2
\times 10^6 cm^{-1}$ in GaAs) and $k_F$ is the Fermi
wavevector.  This expression[18] is the dashed line labeled GQ in
Fig. 3.  Fukuyama and Abrahams (FA)[19], however, found a $T^2
\log(T_F /T)$ term $\pi^2$ larger than GQ; ; this result is
also shown in the figure. Additionally, Fasol[20] and Yacoby,
{\it et al.} [1] both report numerical calculations of $\hbar / \tau_{ee}$
which exceed Eq. 3 by a factor of two.

The measured scattering rate in this experiment is roughly 6X
larger than the GQ result.
We note that Yacoby, {\it et al.} [1] and Berk, {\it et al.} [21],
using different methods, have also reported e-e scattering
rates significantly larger than the GQ prediction. Since our data
possess the temperature and density dependences expected for e-e
scattering in clean 2D systems, and the magnitude is comparable to
the various theoretical estimates, we believe that the operative
broadening mechanism has been identified.  Nevertheless, we comment
on three potential sources of enhanced scattering.  First, in a
{\it diffusive}  2D system the e-e scattering rate is enhanced[22],
with ${\tau_{ee}}^{-1}$ varying linearly with temperature. This
effect, however, is unlikely to be important here. Aside from
possessing a different temperature dependence, the observed
inelastic scattering rate $( \Gamma - \Gamma_0 )/ \hbar$ dominates
the static disorder rate $\Gamma_0 / \hbar$ at high temperature.
This implies that the electrons move ballistically between
inelastic events. Another possible contribution to $\Gamma$ arises
from {\it interlayer}  Coulomb interactions. Recent studies[23] of
the interlayer e-e scattering rate in samples similar to ours show
that this process can be  safely ignored, even after allowing for
the contribution of acoustic interlayer plasmons[24]. In any event,
these interlayer processes cannot be very important since we find
no dependence of $\Gamma$ on the tunnel barrier thickness (between
175 and $340 \AA$).  Finally, we note that Eq. 3 is applicable
only at very low temperatures for electrons exactly at the Fermi
level and assumes that all many-body effects beyond RPA are
negligible.  Each of these assumptions is violated to some extent
by the conditions of our experiment.

In summary, we have applied the technique of 2D-2D tunneling
spectroscopy to determining the spectral properties of electrons in
semiconductor quantum wells.  The density and temperature
dependences of the thermal electron-electron scattering rate have
been measured. We believe that this method will find numerous further
applications.

It is a pleasure to acknowledge fruitful conversations with
E. Abrahams, S. Das Sarma, S.M. Girvin, G.F. Giuliani, P. Hawrylak,
M.S. Hybertsen, T. Jungwirth, A.H. MacDonald, A. Pasquarello, and
Ady Stern.

%\begin{narrowtext}
\begin{figure}
\caption{Typical 2D-2D tunneling resonances observed at
various temperatures in a sample with equal densities ($N _ s
=1.6 \times 10 ^ {11} cm ^ {-2}$) in the two 2DES's.  Insets show
simplified band diagrams on and off resonance.}
\label{autonum}
\end{figure}

\begin{figure}
\caption{Temperature dependence of the tunneling linewidth
$\Gamma$ for three samples.  Samples A and B have comparable 2DES
densities ($N _ s =1.6$ and $1.5 \times 10 ^ {11} cm ^ {-2}$) but
different amounts of static disorder (i.e. different $\Gamma
(T=0)$).  Sample C has a lower density ($0.8 \times 10 ^ {11} cm ^
{-2}$).  Inset: Dotted curves are the ratios $F(V)=I/V$ determined from
the measured $dI/dV$ traces at T=0.74 and 9.1K shown in Fig. 1. (The
point density has been reduced for clarity.)  The definition of the
linewidth $\Gamma$ is shown. The solid curves are Lorentzian fits to
the $F(V)$ data.}
\end{figure}

\begin{figure}
\caption{Tunnel resonance width {\it vs.} temperature for
all samples (having eight different densities).  On dividing $ T $ by
$T _ F$ and the resonance width (minus the zero temperature limit
$\Gamma _ 0$) by $E _ F$ all the data collapse onto a single
curve. The dashed lines are the calculations of Giuliani and Quinn
(GQ)[17] and Fukuyama and Abrahams (FA)[19].  The solid line is
$6.3 \times GQ$. Inset:  Coefficient of $T ^ 2$ term in $\Gamma$
{\it vs.} inverse  density $N _ s ^ {-1}$ (in units of $10 ^ {-11}
cm ^ 2$).}
\end{figure}
%\end{narrowtext}
%\end{narrowtext}

\begin{references}
\bibitem[*]{byline} Present address: Dept. of Physics and Astronomy,
University of Oklahoma, Norman, OK 73019.
\bibitem{yacoby} A. Yacoby, U. Sivan, C.P. Umbach, and J.M. Hong,
Phys. Rev. Lett. {\bf 66}, 1938 (1991); A. Yacoby, M. Heiblum,
H. Shtrikman, V. Umansky, and D. Mahalu, Semicon. Sci. and
Technol. {\bf 9}, 907 (1994).
\bibitem{stern} J. Imry and A. Stern, Semicond. Sci. Technol. {\bf 9}, 1879
(1994).
\bibitem{gornik} The first 2D-2D tunneling experiments were reported by J.
Smoliner,
E. Gornik, and G. Weimann, Appl. Phys. Lett. {\bf 52}, 2136
(1988). See also W. Demmerle, {\it et al.}, Phys. Rev. B {\bf 44},
3090 (1991).
\bibitem{smoliner} J. Smoliner, {\it et al.}, Phys. Rev. B {\bf 47}, 3760
(1993).
\bibitem{sivan}U. Sivan, M. Heiblum, and C.P. Umbach, Phys. Rev. Lett. {\bf
63},
992 (1989).
\bibitem{deltasas}
This neglects the symmetric-antisymmetric splitting $\Delta_{SAS}$ of the
double well structure.
In our samples $\Delta_{SAS} \sim 10 \mu eV$ and is thus ignored.
\bibitem{zheng}
Lian Zheng and A.H. MacDonald, Phys. Rev. B {\bf 47}, 10619 (1993).
\bibitem{technique}
J.P. Eisenstein, L.N. Pfeiffer and K.W. West, Appl. Phys. Lett.,
{\bf 57}, 2324 (1990).
\bibitem{jpe}
Such momentum conservation has also been directly
observed through studies of the
tunneling conductance with a magnetic field applied parallel to the
2D planes. J.P. Eisenstein, T.J. Gramila, L.N. Pfeiffer, and K.W. West,
Phys. Rev. B {\bf 44}, 6511 (1991).
\bibitem{correct}
In those instances where the tunnel current $I$ was directly
measured it agreed excellently with the numerically integrated
$dI/dV$.
\bibitem{highT}
Above about T=2K the thermal acoustic phonon wavevector exceeds $2k_
F$ and large angle scattering dominates $\tau_{eph}$.
For $T\geq 4K$ the mobility becomes significantly temperature dependent
owing to these e-ph processes.
\bibitem{wolf}
See {\it Principles of Electron Tunneling Spectroscopy} by
E.L. Wolf, (Oxford University Press, New York, 1985).
\bibitem{jalabert}
R. Jalabert and S. Das Sarma (Phys. Rev. B{\bf 40},
9723 (1989)) show that the main quasiparticle peak contains roughly
60\% of the total spectral weight for $k=k_ F$.  The remainder is
in a broad incoherent background.
\bibitem{edep}
If $\Gamma$ is significantly energy dependent, Eq. 2 must be slightly
modified.
\bibitem{density}
The densities,
and hence the Fermi energies, are unambiguously determined via the
tunneling itself, by observing the quantum oscillations of $dI/dV$
which appear in weak perpendicular magnetic fields.
\bibitem{hodges}
C. Hodges, H. Smith, and J.W. Wilkins, Phys. Rev. B {\bf 4}, 302
(1971).
\bibitem{gq}
G.F. Giuliani and J.J. Quinn, Phys. Rev. B {\bf 26}, 4421 (1982).
\bibitem{log}
The $\log(2q_{TF} /k_ F )$ term in Eq. 3 introduces a slight
"non-universal" density dependence.  The theoretical results in
Fig. 2 incorporate an intermediate value for this term.  The error
so incurred is about $\pm 5\%$.
\bibitem{fa}
H. Fukuyama and E. Abrahams, Phys. Rev. B {\bf 27}, 5976 (1983).
FA did not, however, calculate the
non-logarithmic $T^2$ contributions to $\hbar / \tau_{ee}$ as
GQ did.  Presumably, these omitted terms would further increase
FA's estimate of the scattering rate.
\bibitem{fasol}
G. Fasol, Appl. Phys. Lett. {\bf 59}, 2430 (1991).
\bibitem{others}
Y. Berk, {\it et al.}, preprint.
\bibitem{altshuler}
B. L. Altshuler, A.G. Aronov, and D.E. Khmelnitsky
J. Phys. C {\bf 15}, 7367 (1982).
\bibitem{gramila}
T.J. Gramila, J.P. Eisenstein, A.H. MacDonald, L.N. Pfeiffer, and
K.W. West, Phys. Rev. Lett. {\bf 66}, 1216 (1991).
\bibitem{hawrylak}
P. Hawrylak, Phys. Rev. B {\bf 35}, 3818 (1987); K. Flensberg and
B. Y.-K. Hu, preprint.
\end{references}
\end{document}